\documentclass[twocolumn,showpacs,preprintnumbers,amsmath,amssymb,prb,superscriptaddress]{revtex4-2}
\usepackage{epsfig}
\usepackage{graphicx}
\usepackage{dcolumn}
\usepackage{bm}
\usepackage{color} 
\usepackage{ulem} 

\usepackage{xspace}

\newcommand{\TK}{$T_{\rm K}$\xspace}

\newcommand{\Tc}{$T_c$\xspace}
\newcommand{\EF}{$E_{\rm F}$\xspace}
\newcommand{\CeRhAs}{CeRh$_2$As$_2$\xspace}
\newcommand{\LaRhAs}{LaRh$_2$As$_2$\xspace}
\newcommand{\CeCuSi}{CeCu$_2$Si$_2$\xspace}
\newcommand{\CeNiGe}{CeNi$_2$Ge$_2$\xspace}
\newcommand{\cf}{$c$-$f$\xspace}
\newcommand{\OC}{$\sigma_1(\omega)$\xspace}
\newcommand{\R}{$R(\omega)$\xspace}
\newcommand{\hw}{$\hbar\omega$\xspace}
\newcommand{\DCOG}{$\Delta \langle \omega \rangle$\xspace}
\newcommand{\DSW}{$\Delta SW$\xspace}

\begin{document}

\title{
Optical study on electronic structure of the locally non-centrosymmetric CeRh$_2$As$_2$
}
\author{Shin-ichi Kimura}
\email{kimura@fbs.osaka-u.ac.jp}
\affiliation{FBS and Department of Physics, Osaka University, Suita, Osaka 565-0871, Japan}
\affiliation{Institute for Molecular Science, Okazaki, Aichi 444-8585, Japan}
\author{J\"org Sichelschmidt}
\affiliation{Max Planck Institute for Chemical Physics of Solids, 01187 Dresden, Germany}
\author{Seunghyun Khim}
\affiliation{Max Planck Institute for Chemical Physics of Solids, 01187 Dresden, Germany}
\date{\today}
\begin{abstract}

The electronic structures of the heavy-fermion superconductor \CeRhAs with the local inversion symmetry breaking
and the reference material \LaRhAs have been investigated by using experimental optical conductivity (\OC) spectra and first-principal DFT calculations.
In the low-temperature \OC spectra of \CeRhAs, 
a $4f$--conduction electron hybridization and heavy quasiparticles are clearly indicated by a mid-infrared peak and a narrow Drude peak.
In \LaRhAs, these features are absent in the \OC spectrum, however, it can nicely be reproduced by DFT calculations.
For both compounds, the combination between a local inversion symmetry breaking 
and a large spin-orbit (SO) interaction plays an important role for the electronic structure, 
however, the SO splitting bands could not be resolved in the \OC spectra due to the small SO splitting size.

\end{abstract}

%
\maketitle
%
\section{Introduction}
Materials with non-centrosymmetric crystal structure and crystal surfaces with inversion symmetry breaking 
have recently attracted attention for novel physical properties combined with spin-orbit interaction (SOI)~\cite{Feng2017}.
By using the SOI and the electric field gradient generated by the spatial symmetry breaking, 
a spin-polarized current originating from spin-polarized bands is generated and is regarded to be useful for spintronics applications.
A breaking of local inversion symmetry also generates a toroidal current owing to the Dzyaloshinsky-Moriya interaction~\cite{Crabtree2018} 
and a Cooper pair with a helical spin structure, which produces a superconducting state with a high critical field~\cite{Frigeri2004}. 
Actually, an extremely high upper critical field $H_{c2}$ as high as $5~{\rm T}$ has been observed in the heavy-fermion superconductor CePt$_3$Si (\Tc~$\sim 0.75~{\rm K}$) having a globally non-centrosymmetric crystal structure~\cite{Bauer2004}.

\begin{figure}[t]
\begin{center}
\includegraphics[width=0.30\textwidth]{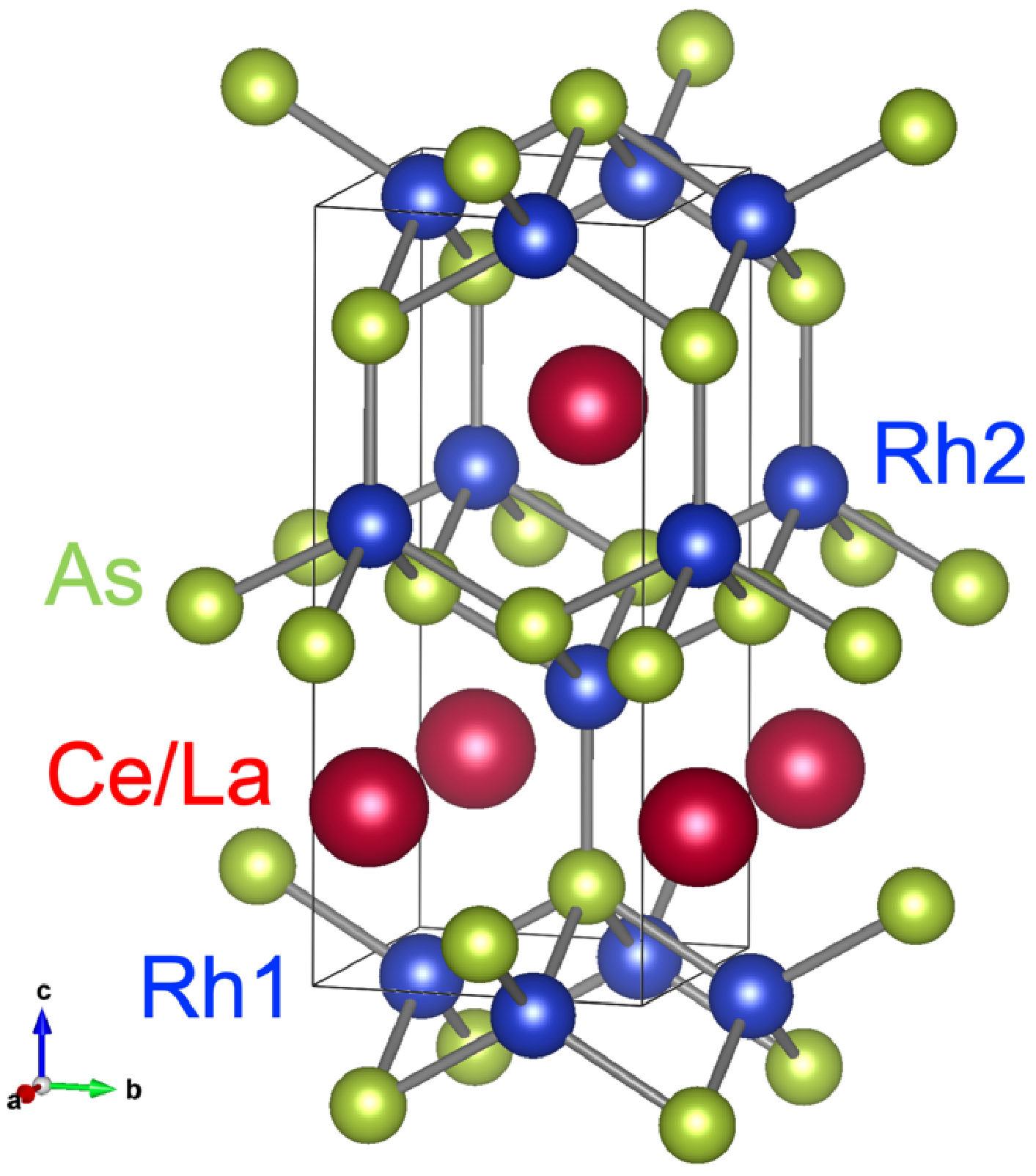}
\end{center}
\caption{
Crystal structure of $R$Rh$_2$As$_2$ ($R=$~Ce, La) with the space group of $P4/nmm$ depited by using VESTA~\cite{Momma2011}. 
Rh~1 and Rh~2 are located at different sites and therefore have different effects on the electronic structure (see Figs.~\ref{fig:LRA_band} and \ref{fig:CRA_band}).
}
\label{fig:crystalstructure}
\end{figure}

\CeRhAs has recently been discovered as a novel heavy-fermion superconductor (\Tc~=~$0.26~{\rm K}$) ~\cite{Khim2021}.
It crystallizes in the CaBe$_2$Ge$_2$-type structure ($P4/nmm$), 
which lacks a local inversion symmetry of the Ce and one of the Rh and As sites 
while maintaining a global inversion center (see Fig.~\ref{fig:crystalstructure}).
Noteworthy, the majority of $RM_2X_2$ ($R$~=~rare earth, $M$~=~transition metal, $X$~=~Si, Ge) compounds 
including the first heavy-fermion superconductor \CeCuSi~\cite{Steglich1979} 
have a tetragonal ThCr$_2$Si$_2$-type crystal structure with inversion symmetry ($I4/mmm$).

The fundamental physical properties of \CeRhAs including superconductivity have recently been reported~\cite{Khim2021}. 
The Kondo temperature (\TK) and the electronic specific heat coefficient ($\gamma$)  
have been evaluated as $20-40$~K and $\approx1$~J/mol K$^2$ ~\cite{Khim2021}, respectively, 
which are similar to those of \CeCuSi~\cite{Reinert2001,Kittaka2014}.
The origin of the superconducting property of locally non-centrosymmetric heavy fermions including \CeRhAs has been investigated theoretically~\cite{Yoshida2015b,Schertenleib2021,Skurativska2021,Nogaki2021,Ptok2021,Cavanagh2021,Mockli2021} and experimentally~\cite{Hafner2021}.
However, the electronic structure of the material, which is the most basic information for the discussion of the physical properties, has not been investigated experimentally so far, although the electronic structure using DFT calculations has been reported~\cite{Nogaki2021,Ptok2021,Cavanagh2021}.

A locally non-centrosymmetric crystal structure can produce a local electric field to the electron orbitals.
With a large SOI, the local electric field induces the band splitting such as the Rashba effect~\cite{Rashba1959,Bihlmayer2015},
which can be directly observed by an angle-resolved photoelectron spectroscopy (ARPES)~\cite{Ishizaka2011} 
and (magneto-optical) infrared spectroscopies~\cite{Demko2012,Martin2013,Martin2016}.
ARPES studies on non-centrosymmetric heavy fermions have been performed on CeIrSi$_3$~\cite{Ohkochi2009} and UIr~\cite{Yamagami2010}.
However, so far, no studies of optical conductivity were reported for this type of materials.

We investigated the fundamental electronic structure of the locally non-centrosymmetric heavy-fermion material \CeRhAs 
by measuring the optical conductivity (\OC) spectra and comparing them with first-principal DFT calculations.
As a reference material without $4f$ electrons, we also investigated \LaRhAs in order to discuss the electronic structure 
without the effect of the hybridization between conduction and $4f$ electrons (\cf hybridization).
Firstly, the obtained \OC spectrum of \LaRhAs is compared with the DFT calculations for the fundamental electronic structure without $4f$ electrons.
Next, the \OC spectrum of \CeRhAs is compared with the corresponding calculation results.
Finally, the temperature-dependent \OC spectra of \CeRhAs is compared with other heavy fermion materials to discuss the evolution of the \cf hybridization.

\section{Experiment and calculation methods}
%
\begin{figure}[t]
\begin{center}
\includegraphics[width=0.45\textwidth]{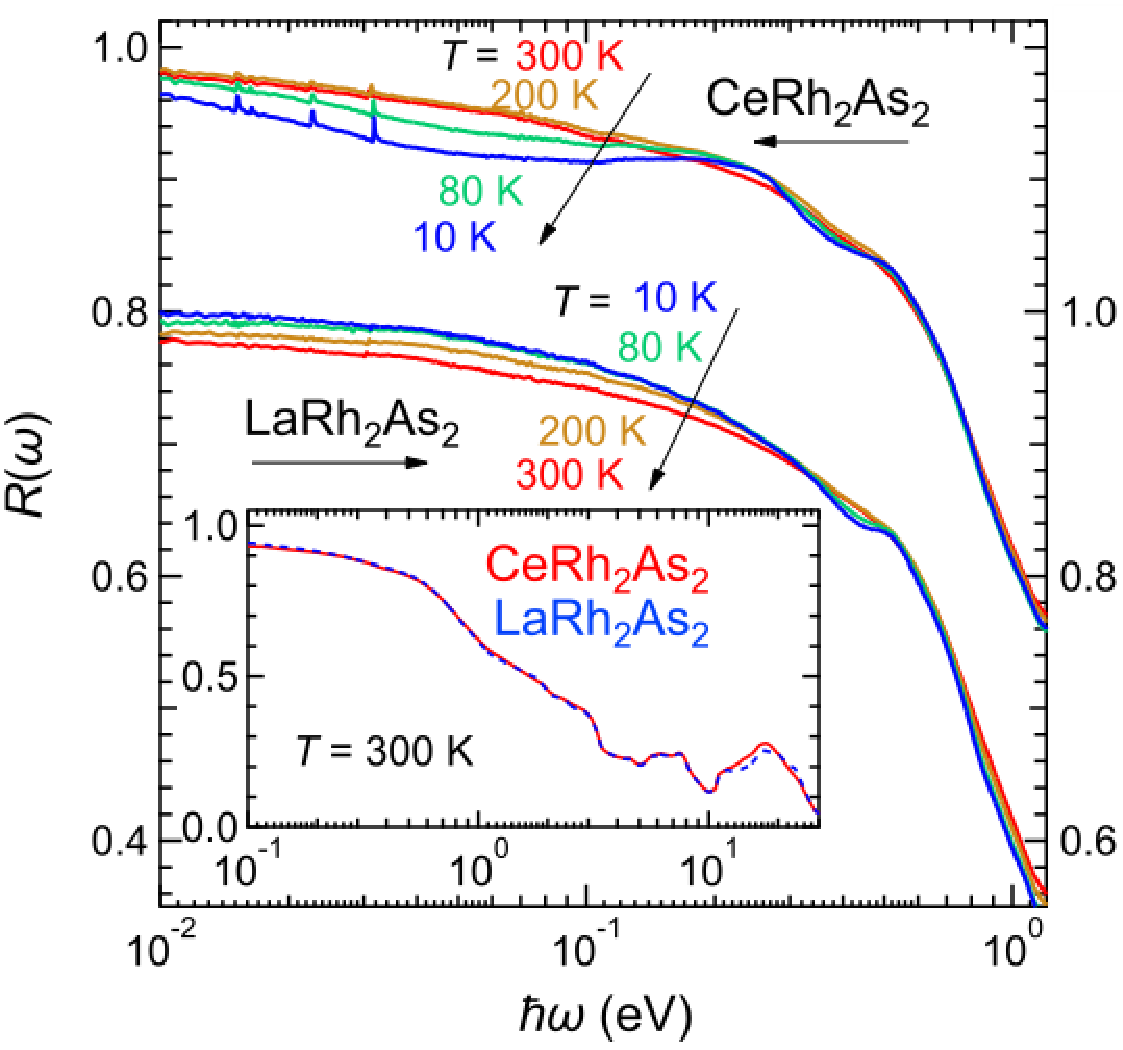}
\end{center}
\caption{
Temperature-dependent reflectivity (\R) spectra of the as-grown $(001)$ surfaces of \CeRhAs and \LaRhAs 
in the photon energy \hw range of $0.01-1.2$~eV.
Fine structures of \CeRhAs at the photon energies \hw of about 15, 23, and 32~meV originate from TO phonons (not discussed here).
(Inset) Wide-energy-range \R spectra of \CeRhAs and \LaRhAs in the \hw range of $0.1-30$~eV at 300~K.
Both spectra are almost identical suggesting the similar electronic structure at room temperature.
}
\label{fig:reflectivity}
\end{figure}

Single-crystalline \CeRhAs and \LaRhAs samples were synthesized by the Bi-flux method~\cite{Khim2021}.
The optical reflectivity \R measurements have been performed using the as-grown (001) plane.
Near-normal-incident \R spectra were acquired in a wide photon-energy range of 8~meV -- 30~eV to ensure accurate Kramers-Kronig analysis (KKA)~\cite{Kimura2013}.
Infrared (IR) and terahertz (THz) measurements at the photon energy \hw regions of 8~meV--1.5~eV have been performed using conventional near-normal reflectivity measurement setups to obtain absolute \R at an accuracy of $\pm0.3$~\% with a feed-back positioning system in the temperature range of 10--300~K~\cite{Kimura2008}.
To obtain the absolute \R values, the {\it in-situ} gold evaporation method was adopted.
Obtained \R spectra of \CeRhAs and \LaRhAs are shown in Fig.~\ref{fig:reflectivity}.
In the photon energy range of 1.5--30~eV, the \R spectrum was measured only at 300~K by using the synchrotron radiation setup at the beamline 3B of UVSOR-III Synchrotron~\cite{Fukui2014}, and connected to the spectra for \hw $\leq 1.5$~eV for KKA.
In order to obtain \OC via KKA of \R, 
the spectra were extrapolated below 8~meV with a Hagen-Rubens function [$R(\omega)=1-\{2\omega/(\pi\sigma_{DC})\}^{1/2}$] due to the metallic \R spectra, 
and above 30~eV with a free-electron approximation $R(\omega) \propto \omega^{-4}$ ~\cite{Dressel2002}.
Here, the values of the direct current conductivity ($\sigma_{DC}$) were adopted from the experimental values \cite{Khim2021}.
The extrapolations were confirmed not to severely affect to the \OC spectra at around 100~meV, which are the main part in this paper.

\begin{table*}[t]

\begin{center}
\caption{
Lattice parameters of \CeRhAs and \LaRhAs used for the DFT calculations.
The parameters were obtained by x-ray diffraction methods.
}

    \begin{tabular}{ c | ccc|ccc}
    \hline
    Sample & & \CeRhAs & & & \LaRhAs & \\ \hline
    Crystal structure & \multicolumn{6}{c}{CaBe$_2$Ge$_2$-type} \\
    Space group & \multicolumn{6}{c}{$P_4/nmm$ (No.~129)} \\ \hline
    Lattice constant & & & & & & \\ 
    a (\AA), b (\AA), c (\AA)  & 4.283(1), & 4.283(1), & 9.865(2) & 4.3137(2), & 4.3137(2), & 9.8803(4) \\
    \hline
    Position & X & Y & Z &X & Y & Z \\ \hline
    Ce, La & 0.25 & 0.25 & 0.254686(3) & 0.25 & 0.25 & 0.2543(2) \\
    Rh 1 & 0.75 & 0.25 & 0 & 0.75 & 0.25 & 0 \\
    Rh 2 & 0.25 & 0.25 & 0.617418(4) & 0.25 & 0.25 & 0.6159(2) \\
    As 1 & 0.75 & 0.25 & 0.5 & 0.75 & 0.25 & 0.5 \\
    As 2 & 0.25 & 0.25 & 0.864075(7) & 0.25 & 0.25 & 0.871(9) \\ \hline
    \end{tabular}

\end{center}

\label{tbl:lattice}

\end{table*}

First-principal DFT calculations have been performed by using the {\sc Wien2k} code including SOI~\cite{Blaha1990} 
to explain the experimentally obtained \OC spectra.
Lattice parameters obtained from room-temperature x-ray diffraction measurements shown in Table~I were adopted to the calculations. 
The obtained band structure of \CeRhAs was consistent with the recent report~\cite{Nogaki2021,Cavanagh2021}.
\OC spectra of the interband transitions have also been calculated by using the {\sc Wien2k} code.
For the discussion of \LaRhAs, we also performed DFT calculations on the basis of a hyphothetical ThCr$_2$Si$_2$-type crystal structure, assuming the same lattice constants as for the original CaBe$_2$Ge$_2$-type structure.

\section{Results and Discussion}

\begin{figure}[t]
\begin{center}
\includegraphics[width=0.40\textwidth]{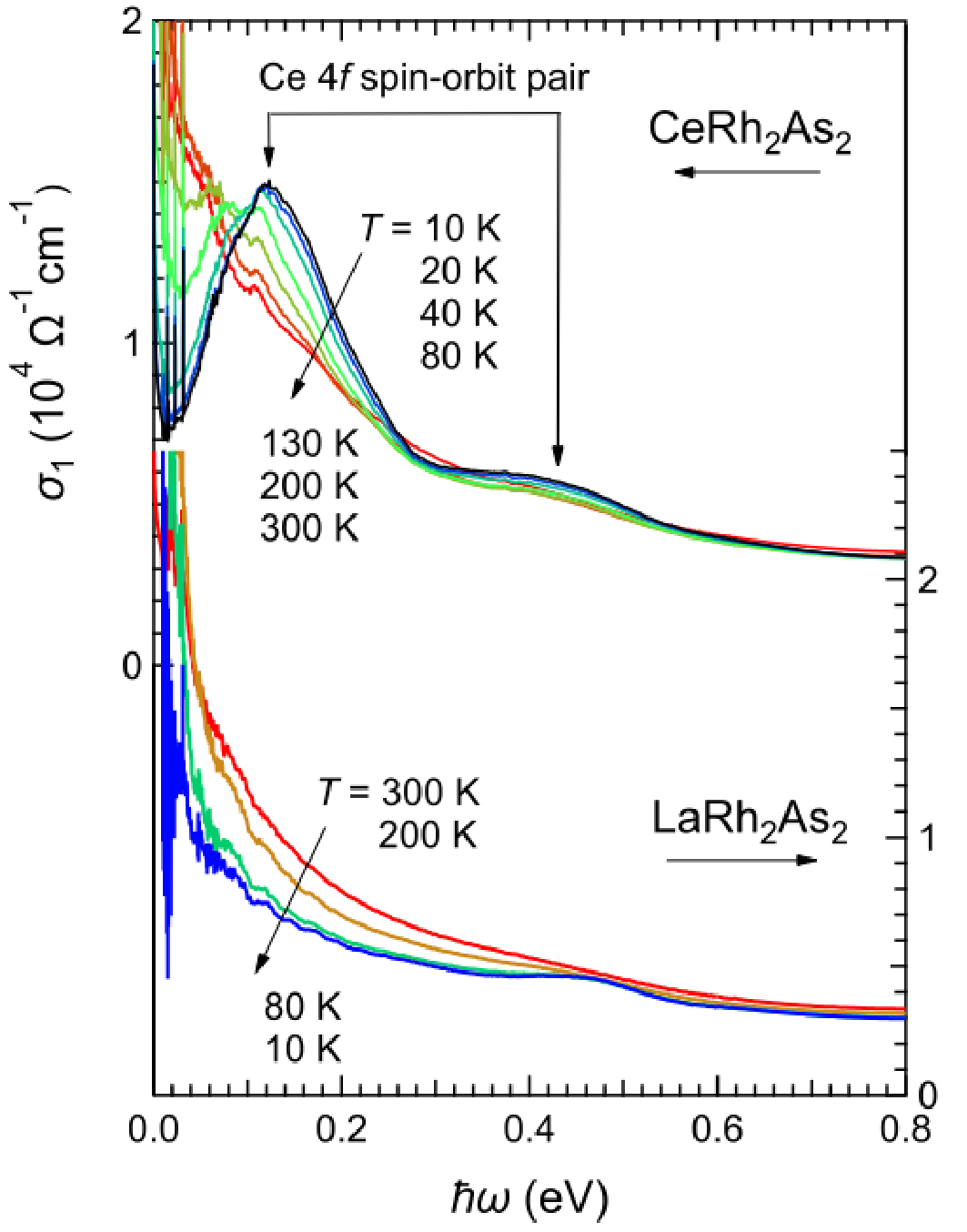}
\end{center}
\caption{
Temperature-dependent optical conductivity (\OC) spectra of \CeRhAs and \LaRhAs.
}
\label{fig:OC}
\end{figure}
The measured \R spectra of \CeRhAs and \LaRhAs, shown in Fig.~\ref{fig:reflectivity}, at various temperatures, 
were used for KKA in order to obtain their \OC spectra as shown in Fig.~\ref{fig:OC}.
At 300~K, \OC of both materials monotonically increase with decreasing \hw suggesting a typical metallic character.
The spectrum of \CeRhAs at 300~K is very similar to that of \LaRhAs, 
as demonstrated by almost identical \R spectra in the inset of Fig.~\ref{fig:reflectivity}. 
This suggests fully localized Ce~$4f$ states at 300~K.
With decreasing temperature, the spectra strongly change and new features emerge:
A double-peak structure (``mid-IR peak'') appears at 0.12 and 0.4~eV in \CeRhAs , 
whereas in \LaRhAs, a weak single peak becomes visible at 0.45~eV.
The mid-IR peak is usually observed in many Ce compounds and 
suggests the emergence of the \cf hybridization~\cite{Kimura2021}.
The peak appearing in \LaRhAs does not obviously originate from the \cf hybridization, 
but can be explained by the electronic structure of \LaRhAs.

\subsection{Electronic structure of \LaRhAs}

\begin{figure*}[t]
\begin{center}
\includegraphics[width=0.8\textwidth]{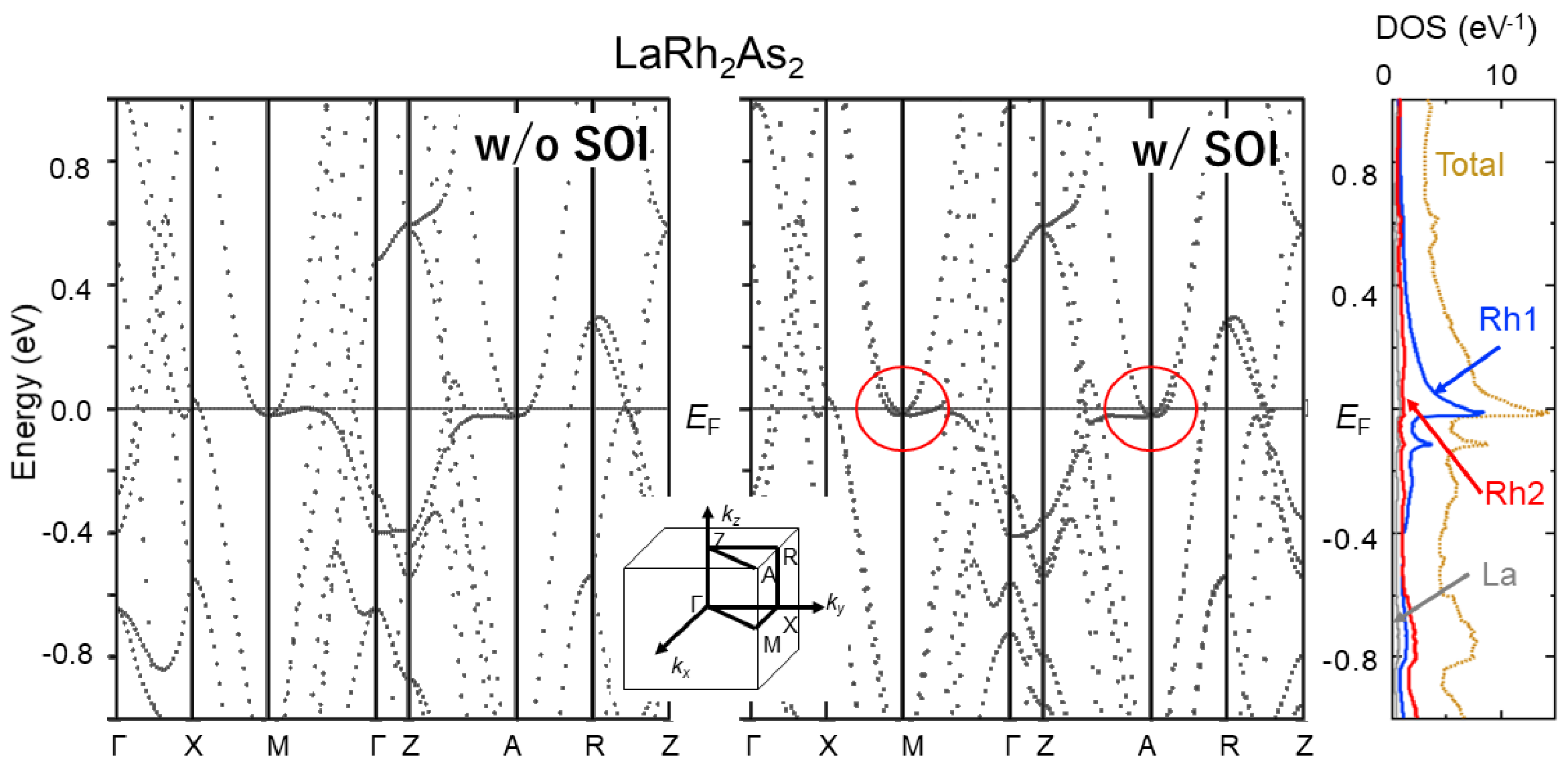}
\end{center}
\caption{
Calculated band structure of \LaRhAs without (left) SOI and with (center) SOI. 
Spin-orbit splitted bands are red encircled.
Inset depicts symmetry points in the Brillouin zone.
(Right) Total density of states and partial density of states of La, Rh~1 and Rh~2 (see Fig.~\ref{fig:crystalstructure}) with SOI.
}
\label{fig:LRA_band}
\end{figure*}

Figure~\ref{fig:LRA_band} shows the band calculation results of \LaRhAs near the Fermi energy (\EF) with and without SOI.
The main effect of including the SOI appears as a spin-orbit (SO) splitting near \EF along the $M-X$ and $A-R$ lines 
(see the red encircled regions in Fig.~\ref{fig:LRA_band}).
The bands mainly originate from the $4d$ states of Rh~1 as shown in the partial density of states depicted in Fig.~\ref{fig:LRA_band}.
The SO splitting is much smaller than that of LaPt$_3$Si~\cite{Uzunok2017} and BiTeI~\cite{Martin2016}.
The different SO splitting size probably originate from the different orbital moments ($4d$ for Rh, $5d$ for Pt, and $6p$ for Bi), 
locally/globally non-centrosymmetric crystal structure,
and/or the different dimensionality, three-dimensional \LaRhAs and LaPt$_3$Si, and two-dimensional BiTeBr.

\begin{figure}[t]
\begin{center}
\includegraphics[width=0.45\textwidth]{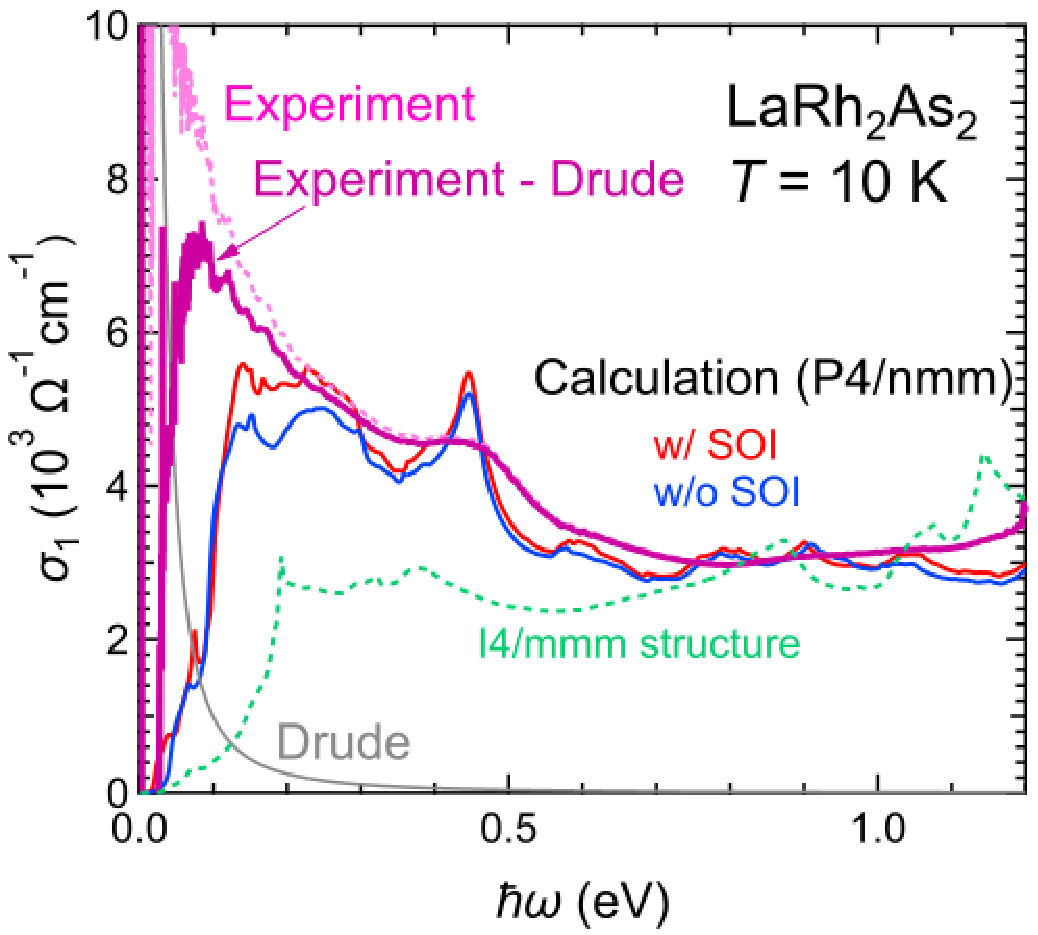}
\end{center}
\caption{
Optical conductivity (\OC) spectrum of \LaRhAs at $T = 10$~K (Experiment, dotted line).
Thick line denotes the interband transition part obtained after subtracting the Drude part (grey line) from the \OC spectrum.
Calculated \OC spectra with and without SOI are shown by red and blue lines, respectively.
The calculated spectrum of a centrosymmetric ThCr$_2$Si$_2$-type crystal structure ($I4/mmm$) is also plotted for comparison (dashed green line).
}
\label{fig:LRA_OC}
\end{figure}

Figure~\ref{fig:LRA_OC} shows the experimental \OC spectrum of \LaRhAs at $T$~=~10~K together with spectra obtained from the DFT calculation results (Fig.~\ref{fig:LRA_band}) either with SOI (red solid line) or without SOI (blue solid line).
The free charge carrier response (Drude peak, grey solid line) evaluated from $\sigma_{DC}$ of \LaRhAs is also shown.
The experimental \OC spectrum (denoted by ``Experiment'' in Fig.~\ref{fig:LRA_OC}) is much larger than the expected Drude curve 
suggesting the existence of other components due to interband transitions overlapping on the Drude curve.
These interband transitions are indicated by the thick solid line (denoted by ``Experiment - Drude'' in Fig.~\ref{fig:LRA_OC}), 
which was derived by subtraction of the Drude curve from the \OC spectrum.
The interband transition spectrum has a broad peak at around $\sim 0.1$~eV and a sharp peak at $\sim 0.5$~eV. 

Both calculated \OC spectra with and without SOI are almost identical suggesting a weak effect of the SOI on \OC. 
This is consistent with the small SOI intensity of \LaRhAs.
The two significant peaks at \hw~$\sim 0.2$~eV and $0.45$~eV can be attributed to 
the experimentally observed peaks at $\sim 0.1$~eV and $\sim 0.5$~eV.
They both originate from the bands near the $\Gamma-M$ and $Z-A$ lines in the Brillouin zone (see Fig.~\ref{fig:LRA_band}).
The background intensity of the calculated spectra is consistent with that of the experimental spectrum. 
Therefore, the experimental \OC spectrum can be explained well by the DFT calculations.

In order to investigate the effect of the crystal structure, Fig.~\ref{fig:LRA_OC} also shows a \OC spectrum calculated by using the centrosymmetric ThCr$_2$Si$_2$-type crystal structure ($I4/mmm$) while keeping the lattice parameters as in the CaBe$_2$Ge$_2$-type structure. 
The large peak at $\sim0.5$~eV does not appear and the background spectral intensity is much lower than that of the experimental spectrum.
This suggests that both the 0.5~eV peak and the high background intensity
are characteristic properties of \LaRhAs and the locally non-centrosymmetric CaBe$_2$Ge$_2$-type crystal structure.

\subsection{Electronic structure of \CeRhAs}

\begin{figure*}[t]
\begin{center}
\includegraphics[width=0.8\textwidth]{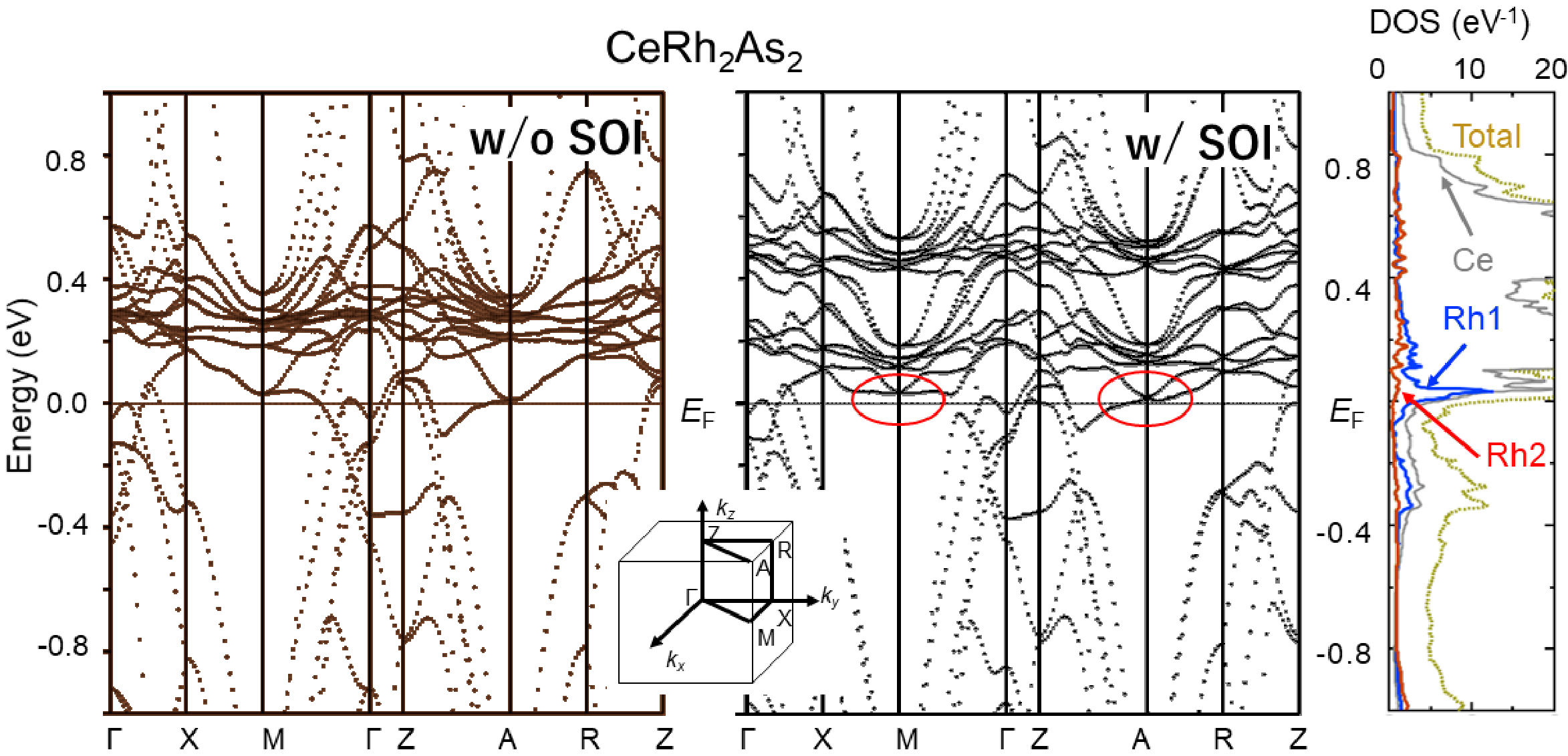}
\end{center}
\caption{
Calculated band structure of \CeRhAs without (left) SOI and with (center) SOI.
Spin-orbit splitted bands are red encircled.
Inset depicts symmetry points in the Brillouin zone.
(Right) Total density of states and partial density of states of Ce, Rh~1 and Rh~2 (see Fig.~\ref{fig:crystalstructure}) with SOI.
}
\label{fig:CRA_band}
\end{figure*}

Figure~\ref{fig:CRA_band} shows the calculated band structures of \CeRhAs with and without SOI.
An important effect of the SOI is a splitting of the Ce~$4f$ bands from almost flat bands at $\sim0.3$~eV into two parts, 
namely $4f_{5/2}$ bands at $\sim0.2$~eV and  $4f_{7/2}$ bands at $\sim0.5$~eV.
The splitting energy is roughly $0.25-0.3$~eV, which is a characteristic value for various Ce-based compounds 
as observed in \OC spectra~\cite{Kimura2021,Kimura2016a} and photoelectron spectra~\cite{Reinert2001,Im2006}.

As in \LaRhAs, the SO split bands appear near \EF along the $M-X$ and $A-R$ lines.
The size of SO splitting is larger than that of \LaRhAs.
Since the bands originate from the hybridization between the Rh~1~$4d$ and Ce~$4f$ bands 
(see the partial densities of states of Ce and Rh~1 in the right figure of Fig.~\ref{fig:CRA_band}), 
the $4f$ state is considered to be relevant for the SO splitting.
Moreover, as flat bands are claimed to be important for the superconductivity~\cite{Nogaki2021}, 
the SO splitting may play an important role for the exotic superconductivity in \CeRhAs.

\begin{figure}[t]
\begin{center}
\includegraphics[width=0.45\textwidth]{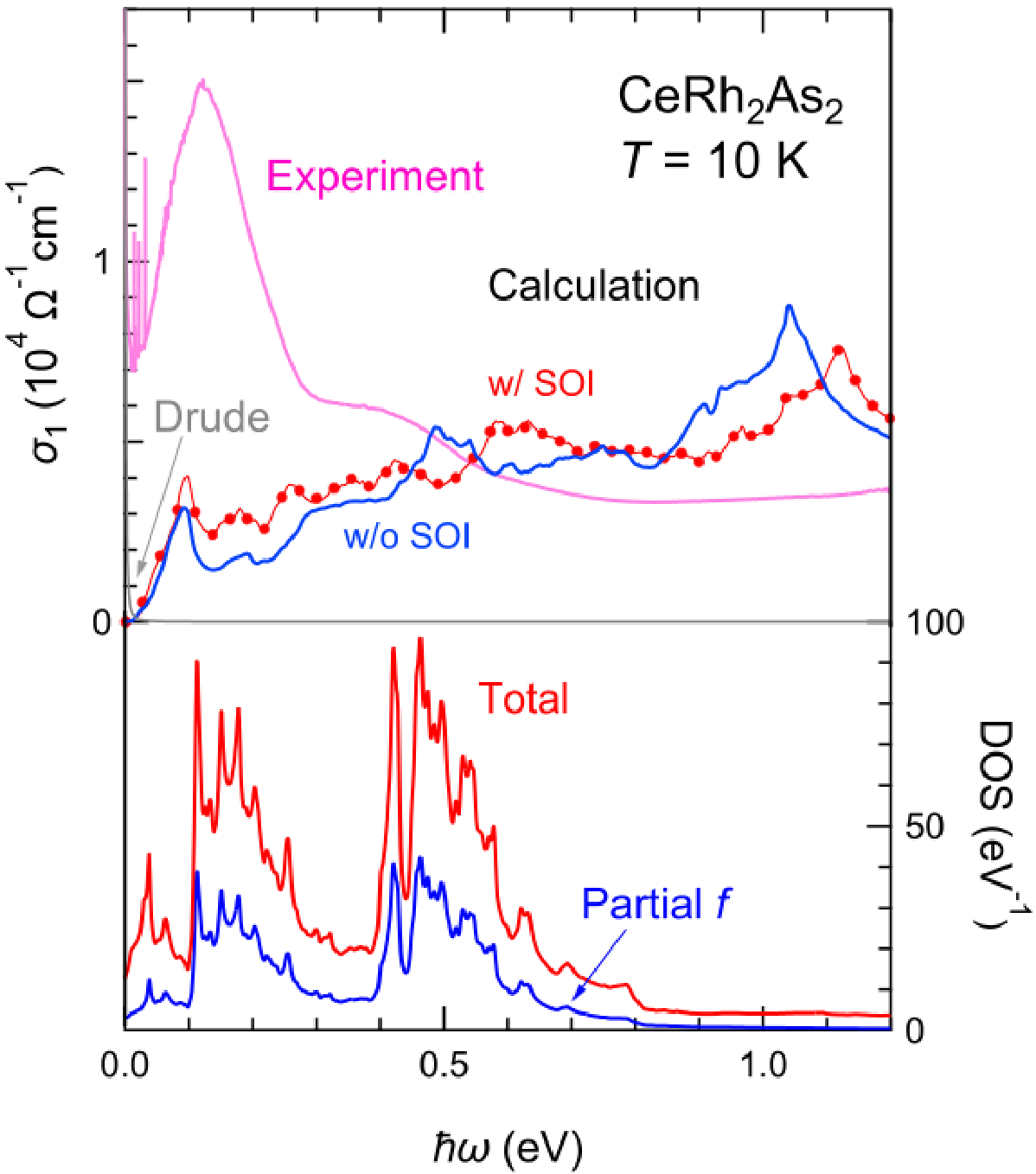}
\end{center}
\caption{
Optical conductivity (\OC) spectrum of \CeRhAs at 10~K (Experiment, magenta solid line).
The Drude curve expected by the experimental electrical resistivity is shown by a gray solid line, 
and calculated \OC spectra with and without SOI are shown by a red marked line and a blue solid line, respectively.
The total and partial $f$ density of states above \EF are also plotted in the lower figure for the comparison with the \OC spectrum.
The \EF of the density of states is set to \hw~$= 0$~eV of \OC spectra.
The peaks at about 0.2 and 0.5~eV originate from the unoccupied Ce~$4f_{5/2}$ and $4f_{7/2}$ states.
}
\label{fig:CRA_OC}
\end{figure}

Figure~\ref{fig:CRA_OC} shows the experimentally obtained \OC spectrum at $T =$~10~K 
together with the calculated \OC spectra with and without SOI. 
In comparison with the \OC spectra of \LaRhAs, the calculated spectra cannot reproduce the experimental spectrum well.
The main reason is that the mid-IR peak below \hw~$\sim 0.6$~eV does not appear in the calculation.
The experimental mid-IR peak can be attributed to the SO splitting of the Ce~$4f$ states~\cite{Kimura2009b}, 
which appears in the unoccupied density of states above \EF as shown in the lower frame of Fig.~\ref{fig:CRA_OC}.
However, the corresponding mid-IR peak structure does not clearly appear in the \OC calculation shown in the upper frame of Fig.~\ref{fig:CRA_OC}.
The inconsistency suggests that the calculated \cf hybridization intensity in the DFT calculations is much smaller than the experimental value 
because the Kondo interaction between conduction and localized $4f$ electrons is not included in the calculation.

According to the band calculation in Fig.~\ref{fig:CRA_band}, 
the signature of the SO splitting along the $\Gamma-M$ line may appear at \hw~$\leq 0.1$~eV.
In this energy region, there are three phonon peaks at \hw~=~15.2, 22.8, and 31.7~meV (see Fig.~\ref{fig:reflectivity}), 
but no other significant structures except for the Drude peak.
Therefore, we conclude that, in the \OC spectra, there are no visible signatures for a SO splitting expected in the band structure.

\subsection{\cf hybridization of \CeRhAs}

\begin{figure}[t]
\begin{center}
\includegraphics[width=0.45\textwidth]{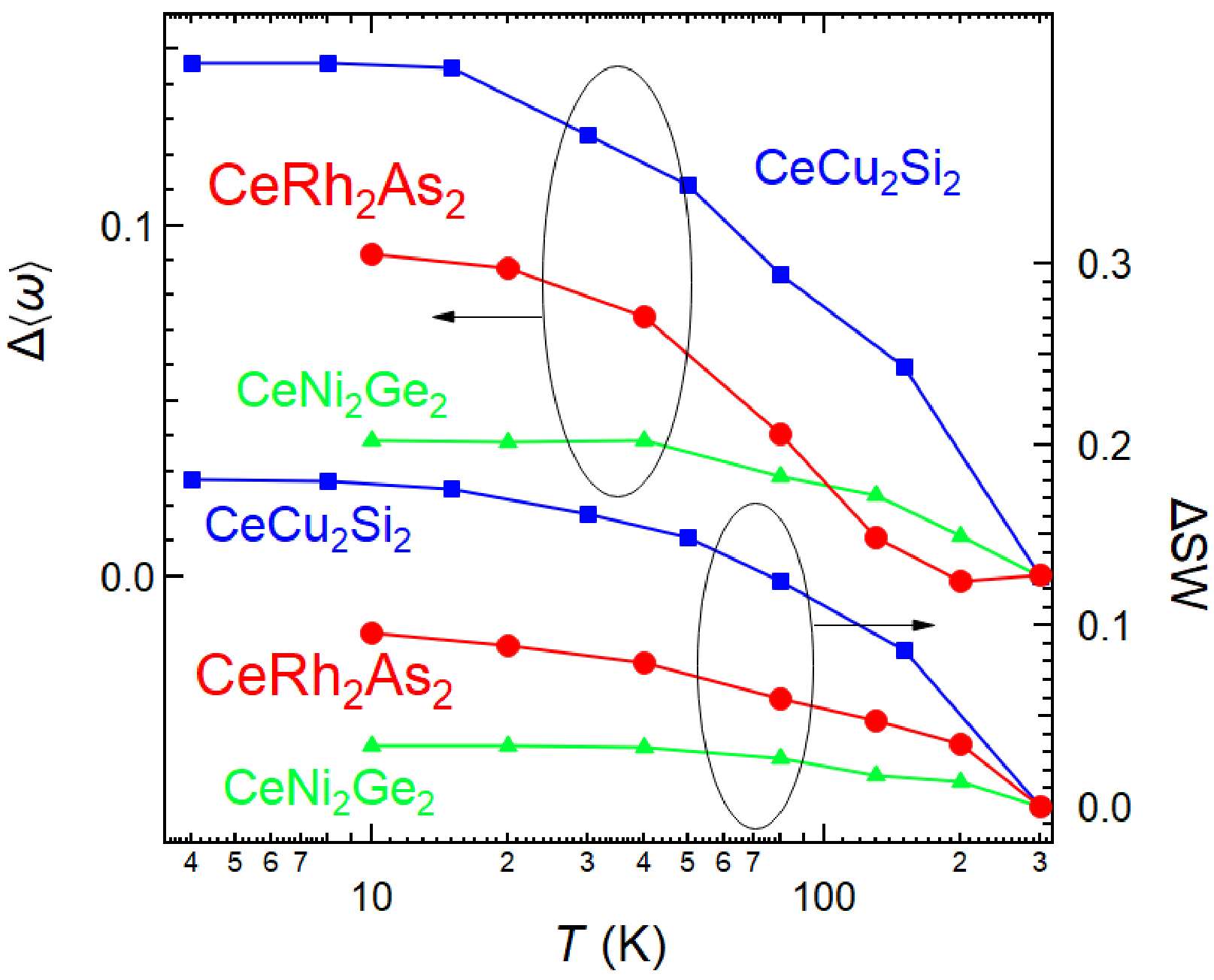}
\end{center}
\caption{
Relative temperature dependence of the change of the center of gravity (\DCOG, left axis) and the spectral weight (\DSW, right axis) 
of \CeRhAs and reference materials, \CeCuSi and \CeNiGe~\cite{Kimura2021}.
These values were normalized to their room-temperature values.
}
\label{fig:Tdep_CRA_OC}
\end{figure}

The \cf hybridization in \CeRhAs can be characterized by using the temperature dependence of the mid-IR peak.
Figure~\ref{fig:Tdep_CRA_OC} shows the temperature dependencies of the center of gravity 
[\DCOG~$=\left[ \langle \omega(T) \rangle - \langle \omega({\rm 300~K}) \rangle \right] / \langle \omega({\rm 300~K}) \rangle$],
where 
$\langle \omega(T) \rangle=\int_{\omega_1}^{\omega_2}\omega\sigma_1(\omega)d\omega / \int_{\omega_1}^{\omega_2}\sigma_1(\omega)d\omega$,
and the spectral weight transfer
[\DSW~$= \int_{\omega_1}^{\omega_2} [|\sigma_1(\omega, T) - \sigma_1(\omega, {\rm 300~K})| / \sigma_1(\omega, {\rm 300~K})] d\omega$] 
of the mid-IR peak of \CeRhAs relative to those at $T=$~300~K.
The integration range was set as $\omega_1 = 0~{\rm eV} \leq \omega \leq \omega_2 = 0.8~{\rm eV}$,
where the spectral change in the lower energy region is almost recovered.
In the figure, the evaluated \DCOG and \DSW of two heavy-fermion compounds \CeCuSi (\TK$\sim10$~K)~\cite{Sichelschmidt2013} and \CeNiGe (\TK$\sim$ several K)~\cite{Kimura2016a} are also plotted for comparison.
The temperature dependencies of \DCOG and \DSW correspond to the evolution of the \cf hybridization with temperature~\cite{Kimura2021}, i.e.,
an increasing itinerancy corresponds to increasing values of \DCOG and \DSW whereas constant values suggest a localization of the $4f$ state.

In \CeRhAs, \DCOG is constant near 300~K and increases toward low temperatures below $\sim150$~K, 
suggesting a rapid development of the \cf hybridization below the temperature.
This is consistent with the \cf hybridization intensity being almost negligible at 300~K, 
which is demonstrated by almost identical \R spectra of \CeRhAs and \LaRhAs at 300~K (see inset of Fig.~\ref{fig:reflectivity}).

The \DSW values monotonically increase with decreasing temperature below 300~K.
As was shown previously~\cite{Kimura2021}, an increase from 300~K to $\sim100$~K originates from electron-phonon interactions, 
and the \cf hybridization effect appears below the temperature of $\sim100$~K.
In \CeRhAs, the \DSW increases at low temperatures, which is consistent with \DCOG.
The values of \DCOG and \DSW of \CeRhAs at the lowest accessible temperature are located 
in between those of \CeCuSi and \CeNiGe suggesting an intermediate \cf hybridization intensity$\tilde{V}$, 
which is included in the Kondo temperature expression 
\begin{equation}
T_{\rm K} \propto \exp[-\tilde{V}^{-2}D_c(E_{\rm F})^{-1}] ,\nonumber
\end{equation}
where $D_c(E_{\rm F})$ is the density of states of the conduction band at \EF~\cite{Hewson1993}.
Hence, compared to \CeCuSi, the higher $T_{\rm K}\sim20-40$~K and smaller $\tilde{V}$ of \CeRhAs 
should be related to a larger $D_c(E_{\rm F})$. 
Indeed, such conclusion is consistent with the \OC spectral intensity of \LaRhAs for \hw$\leq0.8$~eV, 
which reflects the density of states near the \EF. 
In this region, \OC is enhanced compared to the calculated one using the ThCr$_2$Si$_2$-type structure 
as shown in Fig.~\ref{fig:LRA_OC}, which is also due to a flat Rh~$4d$ band near the \EF as shown in Fig.~\ref{fig:LRA_band}. 
Therefore, the locally non-centrosymmetric CaBe$_2$Ge$_2$-type crystal structure of \CeRhAs supports a higher \TK than 
the ThCr$_2$Si$_2$-type structure of \CeCuSi.


\section{Conclusion}
To summarize, optical conductivity \OC spectra of a locally non-centrosymmetric heavy fermion superconductor \CeRhAs and \LaRhAs 
as a reference material without $4f$ electrons were measured and compared with the corresponding DFT calculations.
The experimentally obtained \OC spectrum of \LaRhAs can be explained well by the DFT calculations.
Besides, the experimental \OC spectrum of \CeRhAs at low temperatures has a stronger \cf hybridization intensity than the DFT calculation 
because the Kondo interaction is effective at low temperatures, 
which is also seen in the temperature dependence of the \OC spectrum.
The evidence of the SO splitting due to the locally non-centrosymmetric crystal structure, unfortunately, 
could not be observed in the \OC spectra because of the SO splitting being too small for resolvable spectral features.

\section*{Acknowledgments}
We would like to thank Profs. Noriaki Kimura, Takahiro Ito, Hiroshi Watanabe, and Yoshiyuki Ohtsubo for their fruitful discussion and UVSOR Synchrotron staff members for their support during synchrotron radiation experiments.
Part of this work was performed under the Use-of-UVSOR Synchrotron Facility Program (Proposals No.~20-735) of the Institute for Molecular Science, National Institutes of Natural Sciences.
This work was partly supported by JSPS KAKENHI (Grant No.~20H04453).

%
%

\bibliographystyle{apsrev4-1}
\bibliography{../../../bibtex/library}

\end{document}